\documentclass[aps,twocolumn,superscriptaddress,floatfix]{revtex4}
\usepackage[dvips]{graphicx}
\usepackage{latexsym}
\usepackage{amsmath}
\usepackage{amsfonts}
\usepackage{amssymb}
\usepackage{bm}
\usepackage{color}
\usepackage{txfonts}
\begin{document}
	\newcommand{\fig}[2]{\includegraphics[width=#1]{#2}}
	\newcommand{\la}{{\langle}}
	\newcommand{\ra}{{\rangle}}
	\newcommand{\dg}{{\dagger}}
	\newcommand{\upa}{{\uparrow}}
	\newcommand{\dna}{{\downarrow}}
	\newcommand{\ab}{{\alpha\beta}}
	\newcommand{\ias}{{i\alpha\sigma}}
	\newcommand{\ibs}{{i\beta\sigma}}
	\newcommand{\hH}{\hat{H}}
	\newcommand{\hn}{\hat{n}}
	\newcommand{\hc}{{\hat{\chi}}}
	\newcommand{\hU}{{\hat{U}}}
	\newcommand{\hV}{{\hat{V}}}
	\newcommand{\br}{{\bf r}}
	\newcommand{\bk}{{{\bf k}}}
	\newcommand{\bq}{{{\bf q}}}
	\def\gsim{~\rlap{$>$}{\lower 1.0ex\hbox{$\sim$}}}
	\setlength{\unitlength}{1mm}
	\newcommand{{\vhf}}{$\chi^\text{v}_f$}
	\newcommand{{\vhd}}{$\chi^\text{v}_d$}
	\newcommand{{\vpd}}{$\Delta^\text{v}_d$}
	\newcommand{{\ved}}{$\epsilon^\text{v}_d$}
	\newcommand{{\vved}}{$\varepsilon^\text{v}_d$}
	\newcommand{{\tr}}{{\rm tr}}
	\newcommand{\pprl}{Phys. Rev. Lett. \ }
	\newcommand{\pprb}{Phys. Rev. {B}}

\title {Low-energy effective theory and symmetry classification of  flux phases on Kagome  lattice}
\author{Xilin Feng}
\affiliation{Beijing National Laboratory for Condensed Matter Physics and Institute of Physics,
	Chinese Academy of Sciences, Beijing 100190, China}
\affiliation{School of Physical Sciences, University of Chinese Academy of Sciences, Beijing 100190, China}

\author{Yi Zhang}
\affiliation{Kavli Institute of Theoretical Sciences, University of Chinese Academy of Sciences,
	Beijing, 100190, China}

\author{Kun Jiang}
\email{jiangkun@iphy.ac.cn}
\affiliation{Beijing National Laboratory for Condensed Matter Physics and Institute of Physics,
	Chinese Academy of Sciences, Beijing 100190, China}

\author{Jiangping Hu}
\email{jphu@iphy.ac.cn}
\affiliation{Beijing National Laboratory for Condensed Matter Physics and Institute of Physics,
	Chinese Academy of Sciences, Beijing 100190, China}
\affiliation{Kavli Institute of Theoretical Sciences, University of Chinese Academy of Sciences,
	Beijing, 100190, China}
\date{\today}

\begin{abstract}
Motivated by recent experiments on AV$_3$Sb$_5$ (A=K,Rb,Cs), the chiral flux phase has been proposed to explain time-reversal symmetry breaking. To fully understand the physics behind the chiral flux phase, we construct a low-energy effective theory based on the van-Hove points around the Fermi surface. The possible symmetry-breaking states and their classifications of the low-energy effective theory are completely studied, especially the flux phases on Kagome lattice. In addition, we discuss the relations between the low-energy symmetry breaking orders, the chiral flux  and charge bond  orders. We find all possible 183 flux phases on Kagome lattice within 2*2 unit cell by brute-force approach and classify them by point group symmetry. Among the 183 phases, we find 3 classes in 1*1 unit cell, 8 classes in 1*2 unit cell and
18	classes in 2*2 unit cell, respectively. These results provide a full picture of the time-reversal symmetry breaking in Kagome lattices.
\end{abstract}
\maketitle

In condensed matter physics, there are many interesting unconventional flux phases. For instance, the Haldane model on the honeycomb lattice is the most well-known flux phase, where opposite flux loops are formed in different sublattice triangles respectively \cite{haldane}. 
Meanwhile, flux phases are also widely discussed in high-temperature cuprate superconductors after the seminal work by Affleck and Marston  in t-J models \cite{affleck,lee,doping}. 
Generalizing this discussion, Varma proposed a loop-current phase formed in the Cu-O triangles \cite{varma} and Chakravarty  et al. proposed the d-density wave state with staggered flux in Cu square plaquettes \cite{chakravarty}. 
Both states break the time-reversal symmetry and are supposed to be the candidates for the pseudogap in cuprates \cite{varma06,varma97,keimer,norman}. In addition, flux phases in square lattices, hexagonal lattices and other systems have been widely discussed \cite{Venderbos,yxzhang,nayak,chang,hayami,liu,luo,Shaik}.
Although there are plenty of theoretical proposals, whether flux phases can be found in condensed matter is still an open question.

Recently, the unconventional charge density wave (CDW) order has been found in non-magnetic AV$_3$Sb$_5$ (A=K,Rb,Cs) \cite{ortiz19,ortiz20} by scanning tunneling microscopy (STM) \cite{topocdw} and anomalous hall effect \cite{xhchen,hall20}. This CDW breaks the time-reversal symmetry and is further supported by recent muon spin spectroscopy measurements \cite{topocdw,musr}.
To explain this time-reversal symmetry breaking phenomena, many interesting theoretical proposals have been discussed in Kagome lattice \cite{feng,ronny,nandkishore,balents}, especially the chiral flux phase (CFP) \cite{feng}, which carry unique nontrivial  topological properties.  However, this previous CFP proposal\cite{feng} only includes one particular flux pattern. Several important questions  were left behind, including why the flux pattern is selected, how many flux phases in Kagome lattice and how these flux phases are classified by symmetry. 

In this paper,  we construct  a low-energy effective theory using the dominant scattering between van-Hove (vH) points around Fermi surfaces (FSs) to study the CFPs. We classify the possible symmetry breaking states  by point group operations, including on-site charge orders, bond orders and flux phases. The relations between low-energy breaking orders to the physical orders in real space are completely established, especially the chiral flux phase, charge bond orders proposed in previous works \cite{feng}.  We calculate all kinds of flux configuration in real space within 2*2 unit cell and classify them by symmetry. Our result establishes a full physical picture of the CFPs in AV$_3$Sb$_5$ (A=K,Rb,Cs).
\begin{figure}
	\begin{center}
		\fig{3.4in}{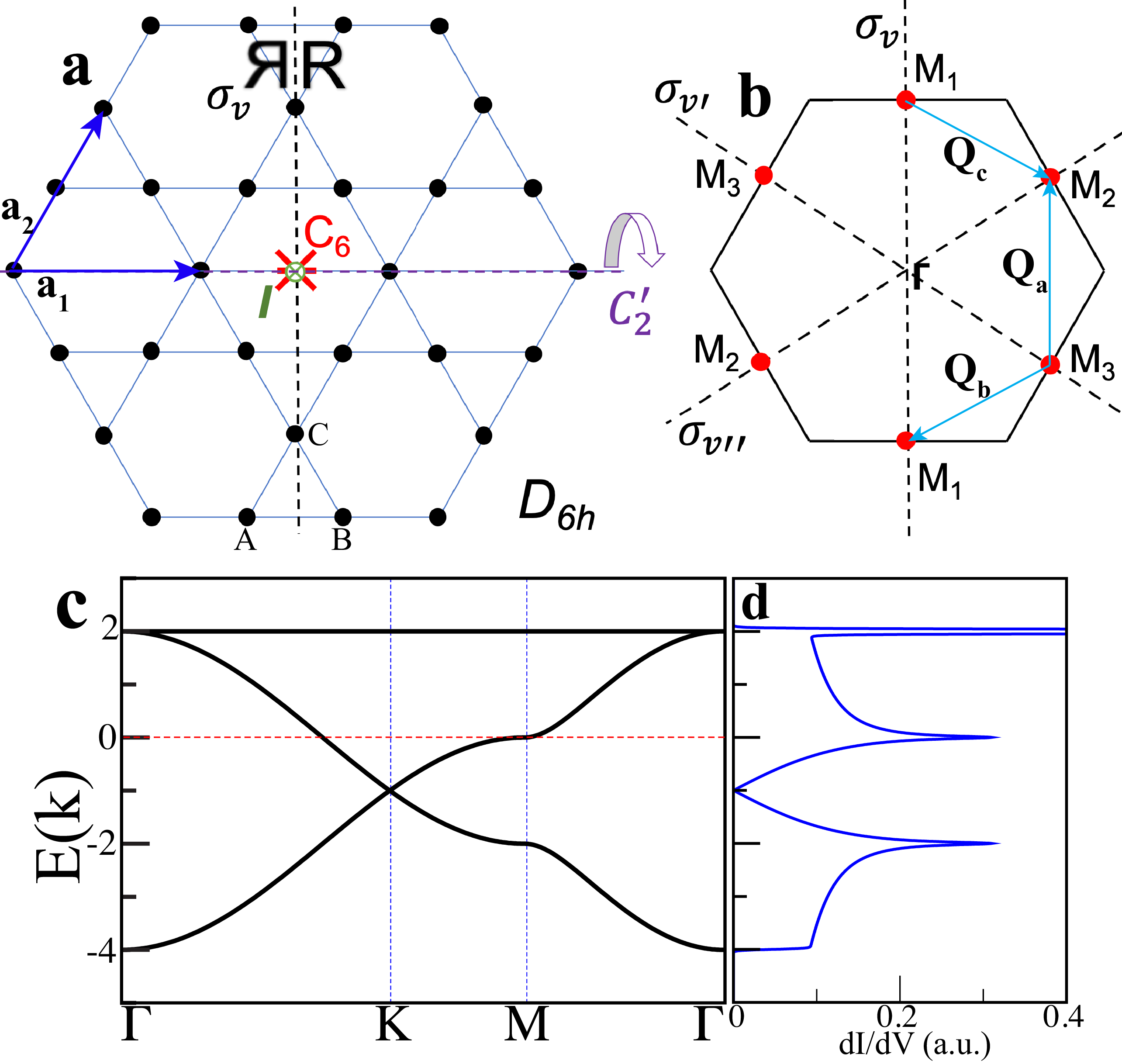}\caption{\textbf{a} Kagome lattice point group $D_{6h}$ and its operations: $C_6$ rotation (red), inversion ${\cal I}$ operation (green), mirror symmetry $\sigma_v$ about the $yz$ plane (black) and $C_2'$ rotation along the x axis (purple). The translation vectors are $\mathbf{a}_1$ and $\mathbf{a}_2$. And the sublattice index is labeled as A,B,C in each unit cell.  \textbf{b} Brillouin zone of Kagome lattice and three dominated van-Hove points M$_1$=$(0,\frac{\pi}{\sqrt{3}})$, M$_2$=$(\frac{\pi}{2},\frac{\pi}{2\sqrt{3}})$ and M$_3$=$(\frac{\pi}{2},-\frac{\pi}{2\sqrt{3}})$.  The corresponding scatter vectors are also labeled as $\mathbf{Q}_a=\{0,\frac{2\pi}{\sqrt{3}}\}$, $\mathbf{Q}_b=\{-\pi,-\frac{\pi}{\sqrt{3}}\}$ and $\mathbf{Q}_c=\{\pi,-\frac{\pi}{\sqrt{3}}\}$. (c) band structure for Kagome lattice. (d) Tunneling density of states (DoS) of tight-binding model for (c), which can be measured by the differential conductance (${d I}/{d V}$).
			\label{BZ}}
	\end{center}
	\vskip-0.5cm
\end{figure}

Before any detailed discussion, we first go through the symmetry group of Kagome lattice, which will be frequently used in the following discussions. The point group of Kagome lattice and AV$_3$Sb$_5$ is $D_{6h}$. The $D_{6h}$ contains three generators: the $C_6$ rotation along the z axis, the inversion operation ${\cal I}$ at the Kagome hexagonal center and the mirror symmetry $\sigma_v$ about the $yz$ plane, as illustrated in Fig.1a. Multiplying $C_6$ rotation generate the $C_3$ and $C_2$ rotations. Multiplying $C_6$, $C_3$ and $C_2$ rotations with ${\cal I}$  generates $S_6$, $S_3$ and $\sigma_h$.  Multiplying $C_6$, $C_3$ and $C_2$ rotations with $\sigma_v$ generate other $\sigma_v$ and $\sigma_d$. Multiplying $\sigma_v$ with ${\cal I}$ generates other $C_2'$s and $C_2''$s. In addition, each unit cell of  Kagome lattice contains three sublattices, labeled as A, B, C, as  shown in Fig.1a. The unit cell forms a triangular lattice with translation vector $\mathbf{a}_1=(1,0)$ and  $\mathbf{a}_2=(\frac{1}{2},\frac{\sqrt{3}}{2})$. This translation group is labeled as $T(\mathbf{a}_1,\mathbf{a}_2)$.  

\section{The low-energy effective theory and 3Q scattering between vH points}
\label{3Q}
As discussed in our previous work \cite{feng,gu,kun}, the electronic properties, especially the CDW orders, of AV$_3$Sb$_5$ materials are dominated by the V $d$ orbitals, which can be captured by a minimum single orbital model \cite{feng,gu,kun}.
To capture the essential physics behind the AV$_3$Sb$_5$  charge density wave, a nearest neighbor tight-binding model on Kagome lattice can be applied without losing generality. 
In the basis of $c_{k}=(c_{k,A},c_{k,B},c_{k,C})$, the Hamiltonian can be written as $H_0=\sum_k c_{k}^\dagger H_k c_{k}$, where
\begin{eqnarray}
	H_k=\begin{bmatrix}
		-\mu & -2t \cos(k_1/2) & -2t \cos(k_2/2) \\
		-2t \cos(k_1/2)  & -\mu & -2t \cos(k_3/2) \\
		-2t \cos(k_2/2) 	&    -2t \cos(k_3/2)  & -\mu
	\end{bmatrix}	
\end{eqnarray}
and  $k_1=k_x$, $k_2=\frac{1}{2}k_x+\frac{\sqrt{3}}{2}k_y$,  and $k_3=-\frac{1}{2}k_x+\frac{\sqrt{3}}{2}k_y$. $\mu$ is the chemical potential and the hopping $t$ is chosen to be 1 as the energy unit.
The band structure for Kagome model is shown in Fig.1c and the electron filling is tuned to the $5/4$ vH filling  ($5/12$ band filling), where the Fermi level crosses the van Hove M points. Throughout this paper,  we use the  Brillouin zone (BZ) filling notation, which is equal to the band filling divided by 3. Owing to the singular density of states (as shown in Fig.1d), the low energy physics of AV$_3$Sb$_5$ should be dominated by the quasiparticles around the vH points. As indicated in the Brillouin zone of Kagome lattice shown in Fig.1b,
there is three vH points at M$_1$=$(0,\frac{2\pi}{\sqrt{3}})$, M$_2$=$(\pi,\frac{\pi}{\sqrt{3}})$ and M$_3$=$(\pi,-\frac{\pi}{\sqrt{3}})$,  The symmetry breaking states of AV$_3$Sb$_5$ are widely believed to come from the scattering between M points with momentum transfer $\mathbf{Q}_a=\{0,\frac{2\pi}{\sqrt{3}}\}$, $\mathbf{Q}_b=\{-\pi,-\frac{\pi}{\sqrt{3}}\}$ and $\mathbf{Q}_c=\{\pi,-\frac{\pi}{\sqrt{3}}\}$.

Hence, we can downfold the model and construct a low-energy effective model based on the  quasiparticles at the three vH points, as 
\begin{equation}
	\psi_M=(\psi_{M_{1}},\psi_{M_{2}},\psi_{M_{3}})^{T}.
\end{equation}
Similar approaches in triangular and honeycomb lattices are discussed in Ref. \cite{Venderbos}.
And for the $5/4$ filling Kagome lattice, the eigenstate of  vH has the exact sublattice index owing to symmetry. Specifically, $\psi_{M_{1}}$ is exactly coming from sublattice $C$, $\psi_{M_{2}}$ is exactly coming from sublattice $A$ and $\psi_{M_{3}}$ is exactly coming from sublattice $B$.

From symmetry point of view, the group of wavevector at M point is $D_{2h}$ and M$_1$, M$_2$, M$_3$ form the star of  M related by the $C_6$ rotation. For simplicity, we choose 5 representative elements  \{$C_{6},C_{3},\sigma_{v},\sigma_{v}',\sigma_{v}''$\} of  $D_{6h}$  to classify the symmetry operations of  $\psi_M$. The three mirror operations $\sigma_{v},\sigma_{v}',\sigma_{v}''$  along the three hexagonal axes are also indicated in Fig.1b.
The matrix elements of each operation in  $\psi_M$ basis are
\begin{widetext}
	\begin{equation}
		\begin{aligned}
			C_6&=&\left(\begin{array}{ccc}0&1&0\\0&0&1\\1&0&0\end{array} \right) ,\   \ \
			C_3&=&\left(\begin{array}{ccc}0&0&1\\1&0&0\\0&1&0\end{array} \right)  ,\   \ \
			\sigma_{v}&=&\left(\begin{array}{ccc}0&0&1\\0&1&0\\1&0&0\end{array} \right)  ,\   \ \
			\sigma_{v'}&=&\left(\begin{array}{ccc}0&1&0\\1&0&0\\0&0&1\end{array} \right)  ,\   \ \
			\sigma_{v''}&=&\left(\begin{array}{ccc}1&0&0\\0&0&1\\0&1&0\end{array} \right) 
		\end{aligned}
	\end{equation}
\end{widetext}	

In 	$\psi_M$ basis,  any symmetry breaking order parameters $\hat{\Delta}_{\alpha}$  can be written as
\begin{eqnarray}
	\hat{\Delta}_{\alpha}=\sum_{i}\Delta_{\alpha, i} \hat{\Gamma}_{i}
\end{eqnarray}
where the $\hat{\Gamma}_{i}$ are the 8 generators  of SU(3) group in the defining representation, which are also known as the Gell-Mann matrices\cite{Venderbos},
\begin{widetext}
\begin{equation}
	\begin{aligned}
		\Gamma_{1}=\left(\begin{array}{ccc}0&1&0\\1&0&0\\0&0&0\end{array} \right),\   \ \Gamma_{2}=\left(\begin{array}{ccc}0&-i&0\\i&0&0\\0&0&0\end{array} \right),\    \ \Gamma_{3}=\left(\begin{array}{ccc}1&0&0\\0&-1&0\\0&0&0\end{array} \right),\     \
		\Gamma_{4}=\left(\begin{array}{ccc}0&0&1\\0&0&0\\1&0&0\end{array} \right), \ \\ \Gamma_{5}=\left(\begin{array}{ccc}0&0&-i\\0&0&0\\i&0&0\end{array} \right),\    
		\Gamma_{6}=\left(\begin{array}{ccc}0&0&0\\0&0&1\\0&1&0\end{array} \right),\     \
		\Gamma_{7}=\left(\begin{array}{ccc}0&0&0\\0&0&-i\\0&i&0\end{array} \right),\     \
		\Gamma_{8}=\frac{1}{\sqrt{3}}\left(\begin{array}{ccc}1&0&0\\0&1&0\\0&0&-2\end{array} \right).
	\end{aligned}
\end{equation}
\end{widetext}
The symmetry properties of  $\hat{\Gamma}_{i}$ under the operations in $D_{6h}$ with $D(R_{i})\Gamma_{i} D(R_{i})^{-1}=D_{ij}\Gamma_{j}$, as summaried in Table.\ref{cosets}.
\begin{table}
	\centering
	\begin{tabular}{c|c|c|c|c|c|c|c|c}
		&$\Gamma_{1}$&$\Gamma_{2}$&$\Gamma_{4}$&$\Gamma_{5}$&$\Gamma_{6}$&$\Gamma_{7}$& $\Gamma_{3}$ & $\Gamma_{8}$ \\ \hline \hline $C_{6}$&$\Gamma_{4}$&$-\Gamma_{5}$&$\Gamma_{6}$&$-\Gamma_{7}$&$\Gamma_{1}$&$\Gamma_{2}$ &$-\frac{1}{2}\Gamma_{3}-\frac{\sqrt{3}}{2}\Gamma_{8}$  & $\frac{\sqrt{3}}{2}\Gamma_{3}-\frac{1}{2}\Gamma_{8}$ \\ \hline $C_{3}$&$\Gamma_{6}$&$\Gamma_{7}$&$\Gamma_{1}$&$-\Gamma_{2}$&$\Gamma_{4}$&$\Gamma_{5}$& $-\frac{1}{2}\Gamma_{3}+\frac{\sqrt{3}}{2}\Gamma_{8}$  & $-\frac{\sqrt{3}}{2}\Gamma_{3}-\frac{1}{2}\Gamma_{8}$ \\ \hline $\sigma_{v}$&$\Gamma_{6}$&$-\Gamma_{7}$&$\Gamma_{4}$&$-\Gamma_{5}$&$\Gamma_{1}$&$-\Gamma_{2}$& $\frac{1}{2}\Gamma_{3}-\frac{\sqrt{3}}{2}\Gamma_{8}$  & $-\frac{\sqrt{3}}{2}\Gamma_{3}-\frac{1}{2}\Gamma_{8}$\\ \hline $\sigma_{v}'$&$\Gamma_{1}$&$-\Gamma_{2}$&$\Gamma_{6}$&$\Gamma_{7}$&$\Gamma_{4}$&$\Gamma_{5}$& $-\Gamma_{3}$ & $\Gamma_{8}$\\ \hline
		$\sigma_{v}''$&$\Gamma_{4}$&$\Gamma_{5}$&$\Gamma_{1}$&$\Gamma_{2}$&$\Gamma_{6}$&$-\Gamma_{7}$& $\frac{1}{2}\Gamma_{3}+\frac{\sqrt{3}}{2}\Gamma_{8}$  & $\frac{\sqrt{3}}{2}\Gamma_{3}-\frac{1}{2}\Gamma_{8}$ 
	\end{tabular}
	\caption{The symmetry relations of $\hat{\Gamma}_{i}$  under the operations in $D_{6h}$.}
	\label{cosets}
\end{table}
From Table.\ref{cosets}, we can find that the  $\hat{\Gamma}_{i}$  can be divided into three classes by operations in $D_{6h}$. They are $\hat{\Delta}_{b}=\{\Gamma_{1},\Gamma_{4},\Gamma_{6}\}$, $\hat{\Delta}_{\phi}=\{\Gamma_{2},\Gamma_{5},\Gamma_{7}\},\hat{\Delta}_{s}=\{\Gamma_{3},\Gamma_{8}\}$.

Hence, the order parameters can be classified based on above transformation relations. For $\hat{\Delta}_{b}$ class, the matrix element for each $\hat{\Delta}_{b}$ is real and gives rise to the inter-scattering between M points. Since $\psi_{Mi}$ carries the sublattice index, the $\hat{\Delta}_{b}$ corresponds to the bonding between sublattices. And the order parameters can be further classified as: 
\begin{eqnarray}
	\hat{\Delta}_{b,1}&=&\Delta_{b,1}(\Gamma_{1}+\Gamma_{4}+\Gamma_{6}) \\
	\hat{\Delta}_{b,2}&=&\Delta_{b,2}(\Gamma_{1}-\Gamma_{4}) \\
	\hat{\Delta}_{b,3}&=&\Delta_{b,3}(\Gamma_{1}-\Gamma_{6})
\end{eqnarray}
where the $\hat{\Delta}_{b,1}$ forms the $A_{1g}$ representation of the point group $D_{6h}$  with breaking the translation symmetry. The $\hat{\Delta}_{b,2}$ and $\hat{\Delta}_{b,3}$ forms the $B_{1g}$ representation of two different $D_{2h}$ groups. These two different $D_{2h}$ groups are generated by three generators: $C_{2}$ rotation along z aixs, $\sigma^{''}_{v}$ or $\sigma_{v}$ for different $D_{2h}$ groups, respectively, inversion operator $\cal I$.

In the same spirit, the $\hat{\Delta}_{\phi}$ class corresponds to flux phase and the order parameters can be classified as: 
\begin{eqnarray}
	\hat{\Delta}_{\phi, 1}&=&\Delta_{\phi, 1}(\Gamma_{2}-\Gamma_{5}+\Gamma_{7}) \\
	\hat{\Delta}_{\phi, 2}&=&\Delta_{\phi, 2}(\Gamma_{5}+\Gamma_{7}) \\
	\hat{\Delta}_{\phi, 3}&=&\Delta_{\phi, 3}(\Gamma_{2}+\Gamma_{5})
\end{eqnarray}
The $\hat{\Delta}_{\phi,1}$ forms the $A_{1g}$ representation of the magnetic point group $D_{6h}^{*}$. This magnetic point group $D_{6h}^{*}$ is normally written as $D_{6h}^{*}(C_{6h})$, where $C_{6h}$ is the invariant subgroup of $D_{6h}$. The $D_{6h}^{*}$ is formed by keeping elements of $D_{6h}$ belonging to the $C_{6h}$ and multiplying the remaining elements by time reversal operator $\cal T$. Hence, the $D_{6h}^{*}$ are generated by three generators: $C_{6}$ rotation along z axis,  inversion operator $I$ and the composite element $\sigma_{v} \cal T$. The first two generators of $D_{6h}^{*}$ generate $C_{6h}$. Later, we will show the low-energy theory of chiral flux phase is exactly $\hat{\Delta}_{\phi,1}$.
The $\hat{\Delta}_{\phi,2}$ and $\hat{\Delta}_{\phi,3}$ belongs to $A_{1g}$ representation of two different $D_{2h}$ groups which are generated by three generators: $C_{2}$ rotation along z axis, inversion operator $\cal I$, $\sigma^{'}_{v}$ or $\sigma^{''}_{v}$ for different $D_{2h}$ groups, respectively.

For the diagonal $\hat{\Delta}_{s}$ class, the order parameters describe the on-site charge difference. $\hat{\Delta}_{s}$ does not involve the scattering between M points and hence does not need to break the translation symmetry. The order parameters can be classified as: 
\begin{eqnarray}
	\hat{\Delta}_{s,1}&=&\Delta_{s,1}(\frac{\sqrt{3}}{2}\Gamma_{3}+\frac{1}{2}\Gamma_{8}) \\
	\hat{\Delta}_{s,2}&=&\Delta_{s,2} \Gamma_{8} 
\end{eqnarray}
The $\hat{\Delta}_{s,1}$, $\hat{\Delta}_{s,2}$ belong to the $A_{1g}$ representations of two different $D_{2h}$ groups.

We can also extend the above discussion to more general multi-orbital cases. As discussed above, the group of wavevector at M point is $D_{2h}$.  $D_{2h}$ only contains 1-dimensional irreducible representations. Hence, $\psi_{M_{i}}$ must belong to one of $D_{2h}$ 1-D  irreducible  representations. We can take $\psi_{M_{1}}$ as an example.  Since $\psi_{M_{1}}$ is the eigenstate of $D_{2h}$ element $\sigma_{v}$, the eigenstate of  $\sigma_{v}$ can be either $C_{C\alpha}$ or  $C_{A\alpha}\pm C_{B\beta}$, where the $\alpha$ and $\beta$ are  corresponding  orbital index. These $\alpha$, $\beta$ orbital should be related to each other by $\sigma_{v}$.  $\psi_{M_{2/3}}$ eigenstates can be found by $C_3$ rotations. 

If the $\psi_{M_{1}}$  is still from C sublattice as  $C_{C\alpha}$, the $\psi_{M_{2}}$  must  be also formed by $C_{A\alpha'}$ and $\psi_{M_{3}}$  is formed by
$C_{B\alpha''}$, where $\alpha'$ and $\alpha''$ are orbitals related by  $C_3$ rotation from $\alpha$ orbital. Hence, the time-reversal breaking flux phase $\hat{\Delta}_\phi$ is still corresponding  to the complex hopping between each sublattice involving the orbital degree of freedom in the Kagome multi-orbital systems, like $C_{C\alpha}^\dagger C_{A\alpha'}$.
On the other hand, if the $\psi_{M_{1}}$  is formed by $C_{A\alpha} \pm C_{B\beta}$. Then, $\psi_{M_{2}}$  is formed by $C_{B\alpha'} \pm C_{C\beta'}$ and $\psi_{M_{3}}$  is formed by 
$C_{A\alpha''} \pm C_{C\beta''}$. The time-reversal breaking flux phase is still dominated by the complex hopping between each sublattice with a partial part of on-site orbital polarization density wave, like  $(C_{A\alpha} \pm C_{B\beta})^\dagger (C_{B\alpha'} \pm C_{C\beta'})$. Besides these cases, any linear combination of $C_{C\alpha}$ and $C_{A\alpha} \pm C_{B\beta}$ is also possible an eigenstate of $\psi_{M_{1}}$, whose flux state is also dominated by the complex hopping.

In short, by using the low-energy model based on the wave functions at three van Hove M points, we discussed the possible symmetry breaking orders. The time-reversal symmetry breaking state dominated by vH points in Kagome lattice must correspond to the complex hopping flux phase between sublattices for both single and multi-orbital models. However, the low-energy model only covers three FS points. The relations between low energy models and the real space pattern of the charge density waves, charge bonds orders and chiral flux phases are still undetermined.
In next section, we will construct the real space order parameters with three $\vec{Q}$ vectors shown in Fig.\ref{BZ} (b), and reveal the relationship between these order parameters and the low energy models constructed in this section.\cite{app}

\section{The relation between low-energy effective model and 3Q pattern in real space}

Besides the low-energy effective model,  the chiral flux phase utilizing the real space 3Q configuration and Kagome sublattice degree of freedom was proposed to be the  reason for time-reversal symmetry breaking in AV$_3$Sb$_5$ \cite{feng}. The key idea is to find a three components vector and each component forms a density wave $\cos(\mathbf{Q}_i\cdot\mathbf{r})$ using one of the three scattering momentum $\mathbf{Q}_i$ between vH points, inspired by previous works in hexagonal lattice vH instabilities \cite{martin,taoli,motome,jxli,qhwang12,qhwang13,thomale12,thomale13,chubukov}.

The first choice is using the charge density for each sublattice as
\begin{eqnarray}
	\hat{\mathbf{n}}(\mathbf{R_{\alpha}})=(\hat{n}_{A_{\alpha}},\hat{n}_{B_{\alpha}},\hat{n}_{C_{\alpha}}).
\end{eqnarray} 
where the $\mathbf{R_{\alpha}}$ is the coordinate for the unit cell formed by the sublattices A, B, C, here, with 2*2 unit cell, it can be divided into four categories: $2n\mathbf{a_{1}}+2m\mathbf{a_{2}}+\{\mathbf{0},\mathbf{a_{1}},\mathbf{a_{2}},\mathbf{a_{1}}+\mathbf{a_{2}}\}$ for $\alpha={1,2,3,4}$, respectively, $n$,$m$ are integers. The $\mathbf{a_{1}}$ and $\mathbf{a_{2}}$ are defined as before. Hence,
the vector charge density wave (vCDW) coupling to this is defined as 
\begin{eqnarray}
	\boldsymbol{\Delta}_{vCDW}(\mathbf{R})=\lambda (\cos(\mathbf{Q}_a \cdot \mathbf{R}), \cos(\mathbf{Q}_b \cdot  \mathbf{R}),\cos(\mathbf{Q}_c  \cdot  \mathbf{R})),
\end{eqnarray}
where $\mathbf{Q}_a=\{0,\frac{2\pi}{\sqrt{3}}\}$, $\mathbf{Q}_b=\{-\pi,-\frac{\pi}{\sqrt{3}}\}$ and $\mathbf{Q}_c=\{\pi,-\frac{\pi}{\sqrt{3}}\}$, as shown in Fig.2(a). Besides this vCDW-a configuration we proposed in the previous work, the other five vCDW configurations (vCDW-b to vCDW-f)  can be also found by permutating the wave momentum $Q_i$ , as shown in Fig.2(b-f).
Among all the vCDW orders, the vCDW-b state has the highest symmetry with the point group $D_{6h}$ and breaking the translation symmetry down to 2*2 order. Moreover, vCDW-b is the lowest energy state among all vCDW orders shown in Fig.2 according to the ground state energy:

\begin{equation}\label{ge}
    E_{g}=<g|\hat{H}|g>,
\end{equation}

where $\hat{H}$ is the mean-field Hamiltonian of vCDW state, and $|g>$ is the corresponding ground state wave function of the mean-field Hamiltonian $\hat{H}$ at 5/4 van Hove filling. For example, the mean-field Hamiltonian of the vCDW-a is:

\begin{equation}
    \hat{H}=H_{0}-\sum_{\mathbf{R_{\alpha}}} \mathbf{\Delta}_{vCDW}(\mathbf{R_{\alpha}}) \cdot \hat{\mathbf{n}}(\mathbf{R_{\alpha}})
\end{equation}

It is worth noting that the charge distribution of vCDW-b state has the same configuration of the chiral flux phase in our previous work. 
Since the chiral flux phase has the magnetic $D_{6h}(C_{6h})$ group symmetry,  the vCDW-b state can coexist with the chiral flux phase and retain its symmetry.
In addition, the symmetry of the vCDW-a and f are both $C_{3h}$ and other remaining three vCDW orders all belong to $C_{2v}$, as shown in Fig2 (c),(d) and (e).
\begin{figure*}
	\begin{center}
		\fig{7.0in}{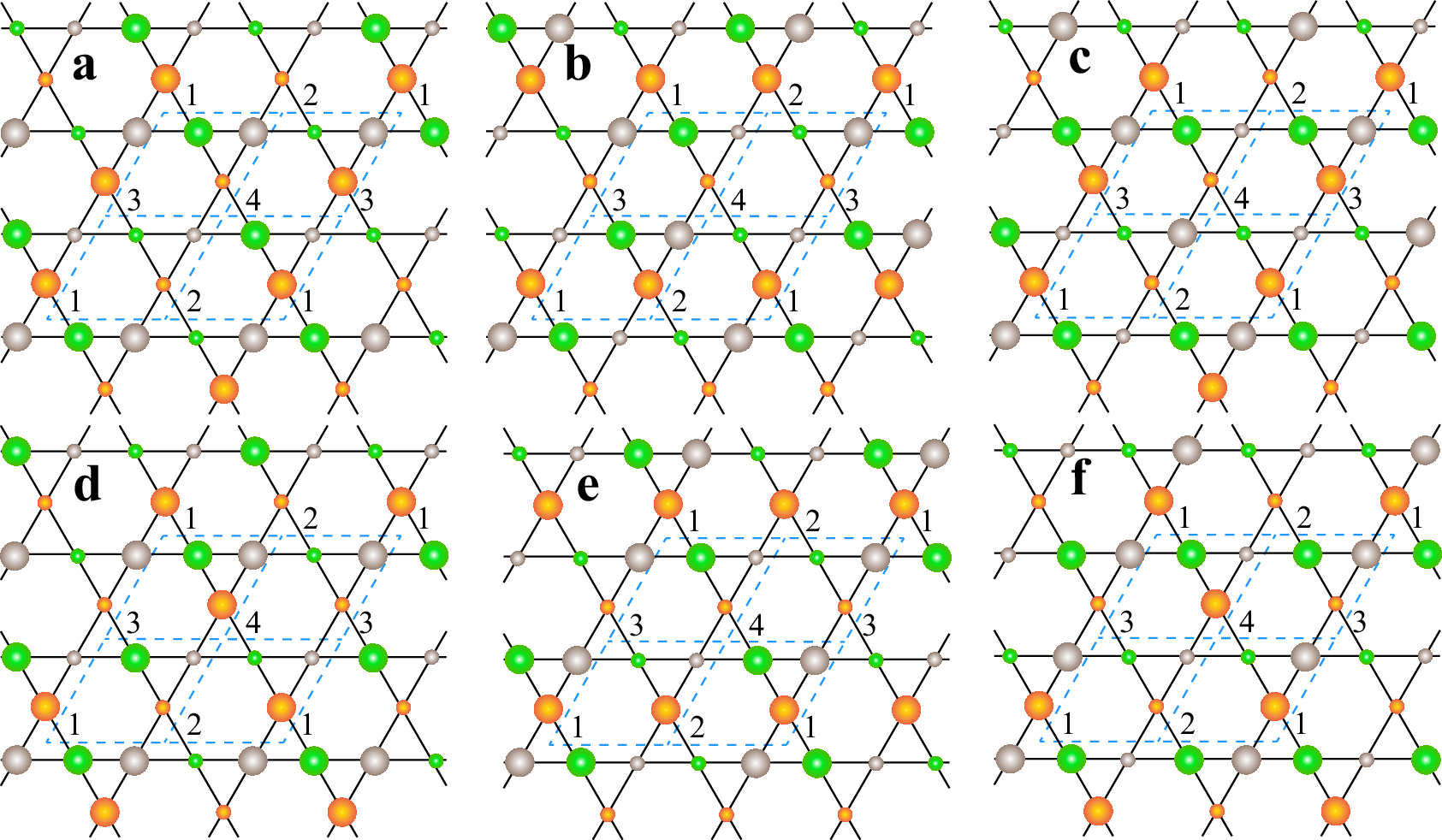}\caption{Six vCDW configurations and their point groups, the size of the dots means the charge density of each site, and different color means different sublattice. (a) vCDW-a ($C_{3h}$),  (b)  vCDW-b ($D_{6h}$), (c)  vCDW-c ($C_{2v}$), (d)  vCDW-d ($C_{2v}$), (e)  vCDW-e ($C_{2v}$),  (f)  vCDW-f ($C_{3h}$).
			\label{vcdw}}
	\end{center}
	\vskip-0.5cm
\end{figure*}

Another choice is  to use the bonds between sublattices
\begin{eqnarray}
	\hat{\mathbf{O}}(\mathbf{R_{\alpha}})=(c_{A_{\alpha}}^\dagger c_{B_{\alpha}},c_{B_{\alpha}}^\dagger c_{C_{\alpha}},c_{C_{\alpha}}^\dagger c_{A_{\alpha}}).
\end{eqnarray}
The charge bond order (CBO) with real order parameter is defined as
\begin{eqnarray}
	\boldsymbol{\Delta}_{CBO}(\mathbf{R})=\lambda (\cos(\mathbf{Q}_a \cdot \mathbf{R}), \cos(\mathbf{Q}_b \cdot  \mathbf{R}),\cos(\mathbf{Q}_c  \cdot  \mathbf{R}))
\end{eqnarray}
as shown in Fig.3(a). Similarly, we can also find other 5 CBO configurations, as shown in Fig.3(b-f).
The symmetry group of the CBO-a state is $D_{3h}$. The CBO-b,c,d orders belong to the $C_{2v}$ group. And the CBO-e,f belong to $C_{3h}$ group. Among all six CBO configurations shown in Fig.3, the CBO-a has the highest symmetry and the lowest energy, which can be gotten by the same method as in vCDW states.
\begin{figure*}
	\begin{center}
		\fig{7.0in}{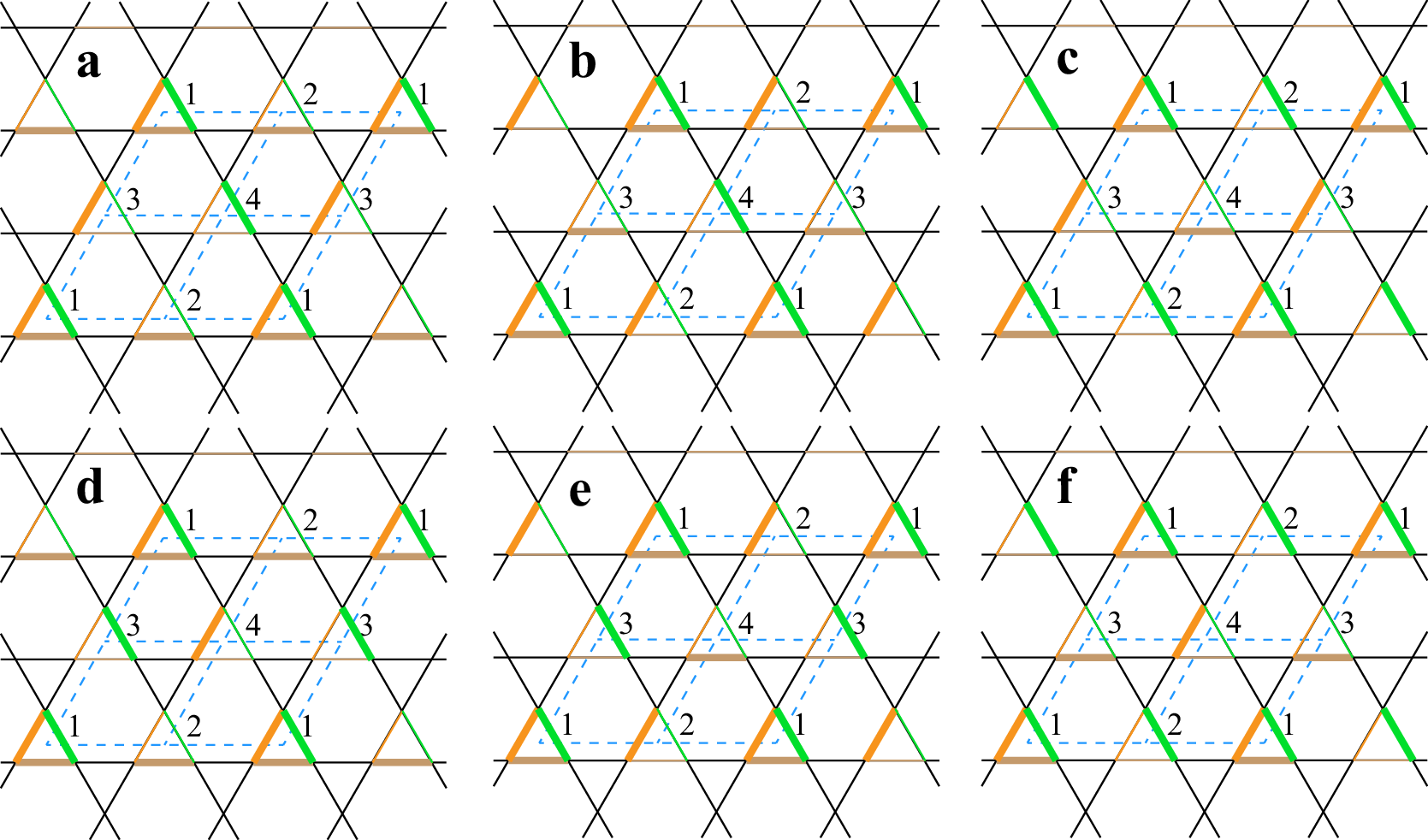}\caption{Six CBO configurations and their point groups, the width of the bond means the relative hopping amplitude, and the color of these bonds mean different hoppings between sublattices. (a) vCBO-a ($D_{3h}$),  (b)  vCBO-b ($C_{2v}$), (c)  vCBO-c ($C_{2v}$), (d)  vCBO-d ($C_{2v}$), (e)  vCBO-e ($C_{3h}$),  (f)  vCBO-f($C_{3h}$).
			\label{cbo1}}
	\end{center}
	\vskip-0.5cm
\end{figure*}

\begin{figure}
	\begin{center}
		\fig{3.4in}{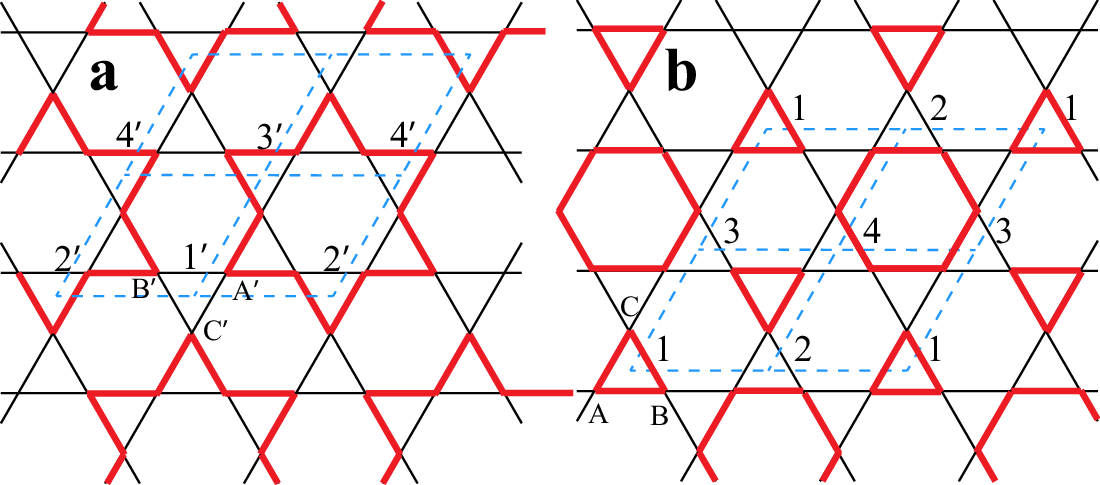}\caption{(a) "Anti-Tri-Hexagonal" bond order  configuration. In order to cover all bonds, we also define another coordinate for Kagome lattice, labeled as $A'$, $B'$, $C'$. To utilize the three Q pattern, the unit cell name of new coordinated is also shifted to $1', 2', 3', 4'$. The red bonds mean the hopping in these bonds are strengthened (b) "Tri-Hexagonal"  bond order configuration.
			\label{cbo2}}
	\end{center}
	\vskip-0.5cm
\end{figure}

Interestingly, another two bond orders "Anti-Tri-Hexagonal" (ATH) and "Tri-Hexagonal" (TrH) in Kagome lattice have been widely discussed \cite{ronny,thomale12,thomale13,balents,wilson,tsirlin}, as shown in Fig.\ref{cbo2}. These two bond orders cover all kagome lattice bonds, which is beyond our above discussion. To include these, the order parameters can be constructed as:
\begin{equation}
\begin{aligned}
    \hat{\mathbf{O}}_{1}(\mathbf{R_{\alpha}})=(c_{A_{\alpha}}^{\dagger}c_{B_{\alpha}},c_{B_{\alpha}}^{\dagger}c_{C_{\alpha}},c_{C_{\alpha}}^{\dagger}c_{A_{\alpha}}),\\
    \hat{\mathbf{O}}_{2}(\mathbf{R'_{\beta}})=(c_{A^{'}_{\beta}}^{\dagger}c_{B^{'}_{\beta}},c_{B^{'}_{\beta}}^{\dagger}c_{C^{'}_{\beta}},c_{C^{'}_{\beta}}^{\dagger}c_{A^{'}_{\beta}}),
\end{aligned}
\end{equation}
where $\mathbf{R_{\alpha}}$ is unit cell coordinate defined above and $\mathbf{R'_{\beta}}$ is the new coordinate for the unit cell formed by the sublattices $A'$, $B'$, $C'$ shown in Fig.\ref{cbo2}(a), and it can be also divided into $2n\mathbf{a_{1}}+2m\mathbf{a_{2}}+\{\mathbf{0},\mathbf{a_{1}},\mathbf{a_{2}},\mathbf{a_{1}}+\mathbf{a_{2}}\}$ for $\beta=1^{'},2^{'},3^{'},4^{'}$, respectively, $n$,$m$ are integers.
The Hamiltonian can be expressed as:
\begin{equation}
H_{CBO}=H_{0}-\sum_{\mathbf{R_{\alpha}}} \mathbf{\Delta}_{CBO}(\mathbf{R_{\alpha}}) \cdot \hat{\mathbf{O}}_{1}(\mathbf{R_{\alpha}})-\sum_{\mathbf{R'_{\beta}}} \mathbf{\Delta}_{CBO}(\mathbf{R'_{\beta}}) \cdot \hat{\mathbf{O}}_{2}(\mathbf{R'_{\beta}}).
\end{equation}
where the $\mathbf{\Delta}_{CBO}(\mathbf{R_{\beta}'})$ is using the same density-wave vectors as in Eq.17. Hence, the ATH and TrH also utilize the three Q scattering mechanism discussed above.
The symmetry group of ATH and TrH is $D_{6h}$. And the low energy effective model of ATH and TrH correspond to  $\hat{\Delta}_{b,1}$, which form the $A_{1g}$ representation of $D_{6h}$ (see Supplement Material (SM) for details\cite{app}). In addition to ATH and TrH, the "Star of David" (SoD) state is another widely proposed configuration \cite{topocdw,yan,wilson}, as shown in Fig.\ref{D6h_CBO}(c). Notice that the SoD state looks quite similar to the ATH bond order, which also belongs to  $D_{6h}$ point group and  $\hat{\Delta}_{b,1}$ effective theory. The TrH state is always the lowest energy state among all the charge bond order states in our calculation. In the next section, we will also discuss the possible CBOs with $C_6$ symmetry.


Finally, if the order parameters coupled to bonds are imaginary, the chiral flux phase can be found 
\begin{eqnarray}
	\boldsymbol{\Delta}_{CFP}(\mathbf{R})=i\lambda (\cos(\mathbf{Q}_a \cdot \mathbf{R}), \cos(\mathbf{Q}_b \cdot  \mathbf{R}),\cos(\mathbf{Q}_c  \cdot  \mathbf{R})).
\end{eqnarray}
 Further adding other terms, the Hamiltonian for CFP can be expressed as:
\begin{equation}
    H_{CFP}=H_{0}-\sum_{\mathbf{R_{\alpha}}} \mathbf{\Delta}_{CFP}(\mathbf{R_{\alpha}}) \cdot \hat{\mathbf{O}}_{1}(\mathbf{R_{\alpha}})-\sum_{\mathbf{R'_{\beta}}} \mathbf{\Delta}_{CFP}(\mathbf{R'_{\beta}}) \cdot \hat{\mathbf{O}}_{2}(\mathbf{R'_{\beta}}).
\end{equation}
The low-energy effective theory of CFP corresponds to $\hat{\Delta}_{\phi,1}$, which belongs to $A_{1g}$ representation of $D_{6h}^{*}(C_{6h})$ magnetic group as discussed in Sec.\ref{3Q}.
Besides this CFP state, we find 122 flux phases in kagome lattice with 2*2 configuration, which will be discussed in the next section.




\section{All possible flux phases in Kagome lattice within 2*2 unit cell}
The above discussion focuses on the low-energy scattering between vH points, which leads to the promising states, chiral flux phase. Are there other flux phases in Kagome lattice? To answer this question, we should find a general principle.
 Generally speaking, a current operator from site $j$ to site $i$ can be found to be
\begin{eqnarray}
	\hat{\mathbf{J}}_{ij}=\frac{e}{i\hbar} \{ t_{ij}C_{i}^\dagger C_j-t_{ij}^{*}C_{j}^\dagger C_i\}
\end{eqnarray}
where $t_{ij}$ is the hopping parameter from site $j$ to site $i$.
Therefore, the expectation value of current operator can only be finite when $t_{ij}$ contains an imaginary part, which corresponds to the flux state.
For any current state, the charge continuity equation $\frac{\partial \rho}{\partial t}+\nabla \cdot \mathbf{J}=0$ must be satisfied, where $\rho$ is the charge density. Therefore,
to find all possible flux solutions, there is only one principle: the currents must conserve at each lattice site without generating any charge sink or source. To simplify our discussion, we also assume that  the amplitudes of the complex hopping terms at all bonds must be the same. Generally speaking, the physical system always favors the high symmetry state. The equal number of arrows going in and out is enough for general cases, such as the Haldane model\cite{haldane}, the loop-current model \cite{varma}, and d-density wave model \cite{chakravarty}.
Following the above constraints, we find 183 flux phases in Kagome lattice within the 2*2 unit cell. Specifically, there are 10 flux phases in 1*1 unit cell, 122 phases in 2*2 unit cell and the remaining 17$\times$3 in 1*2 unit cell, which is discussed separately in the following sub-sections. 

\subsection{1*1 configuration}
\begin{figure*}
	\begin{center}
		\fig{7.0in}{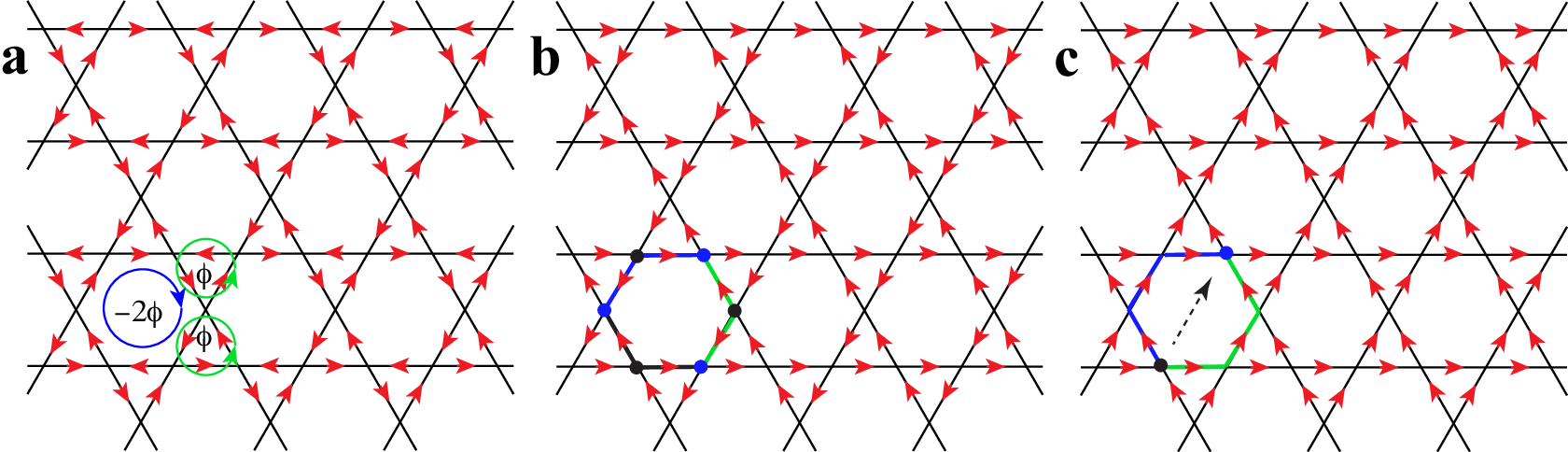}\caption{1*1 flux configurations \textbf{a} Nagaosa solution with $-2\phi$ flux at each hexagon and $\phi$ flux at each triangle. \textbf{b} Flow-b solution with three charge sink (blue dot) and three charge source (black dot) at each hexagon. \textbf{c} Flow-b solution with one charge sink (blue dot) and one charge source (black dot) at each hexagon. The dashed black arrow indicates the flow direction at each hexagon.
			\label{1*1}}
	\end{center}
	\vskip-0.5cm
\end{figure*}

The first flux phase in Kagome lattice is initial proposed by Ohgushi, Murakami and Nagaosa \cite{nagaosa}, which mapped a spin itinerant system with non-zero spin chirality to a flux phase in Kagome lattice. We name this phase as the Nagaosa solution, as shown in Fig.\ref{1*1}a. At each Kagome triangle, there is one flux $\phi$ owing to the complex hopping along the triangle loop. Another flux with $-2\phi$ penetrates each Kagome hexagonal plaquette. The quantum anomalous Hall effect is also obtained in this Nagaosa solution \cite{nagaosa}.
By reversing the current direction, a second Nagaosa  solution can also be obtained.

We can also understand the Nagaosa solution from another point of view. As shown in Fig.\ref{1*1}, the Kagome lattice contains three kinds of bond directions. For the Nagaosa solution, the current direction alternates in each bond direction. Therefore, if the currents all flow in the same direction, the other 1*1 flux configurations can be found. Among these $2^3$  configurations, we also find two classes named flow-a solution and flow-b solution respectively.  For flow-a solution, there are three charge sinks and three charge sources at each hexagon, as shown in Fig.\ref{1*1}b. Owing to charge conservation, the sinks become sources at the neighboring hexagon. Flow-a class contains 2 configurations.  On the other hand, there is one sink and one source at each hexagonal plaquette diagonal direction for flow-b solution, as shown in Fig.\ref{1*1}c. The net flow direction at each hexagon is also labeled as a dashed black arrow.
Since there are also 3 diagonal directions, the flow-b class contains $2\times 3$  configurations.  

In all of these three classes, Nagaosa solution and flow-a solution only break time reversal symmetry and preserve all the point-group symmetry of Kagome lattice. Their symmetry can be described by magnetic group $D_{6h}^{*}$. Since the invariant subgroup of $D_{6h}$ can be either $C_{6h}$ or $D_{3h}$, there are two kinds of magnetic groups. The symmetry of Nagaosa solution can be described by group $D_{6h}^{*}(C_{6h})$, while the symmetry of flow-a solution can be described by $D_{6h}^{*2}(D_{3h})$. Additionally, the flow-b solution breaks both time reversal symmetry and point-group symmetry, which belongs to $D_{2h}^{*}(C_{2v})$ magnetic group.

\begin{table}[ht]
	\begin{ruledtabular}
		\caption{3 classes of flux phase in 1*1 unit cell. The number of configurations at each class is also listed at their brackets.}\label{tab2}
		\begin{tabular}{c|cccc}
			Symmetry & Class Name \\\hline
			$D_{6h}^{*}(C_{6h})$ & Nagaosa (2)                       \\ \hline
			$D_{6h}^{*2}(D_{3h})$ & Flow-a  (2)                           \\ \hline
			$D_{2h}^{*}(C_{2v})$          & Flow-b (6) \\ 
		\end{tabular}
	\end{ruledtabular}
\end{table}

\subsection{2*2 configuration}
\begin{figure}
	\begin{center}
		\fig{3.4in}{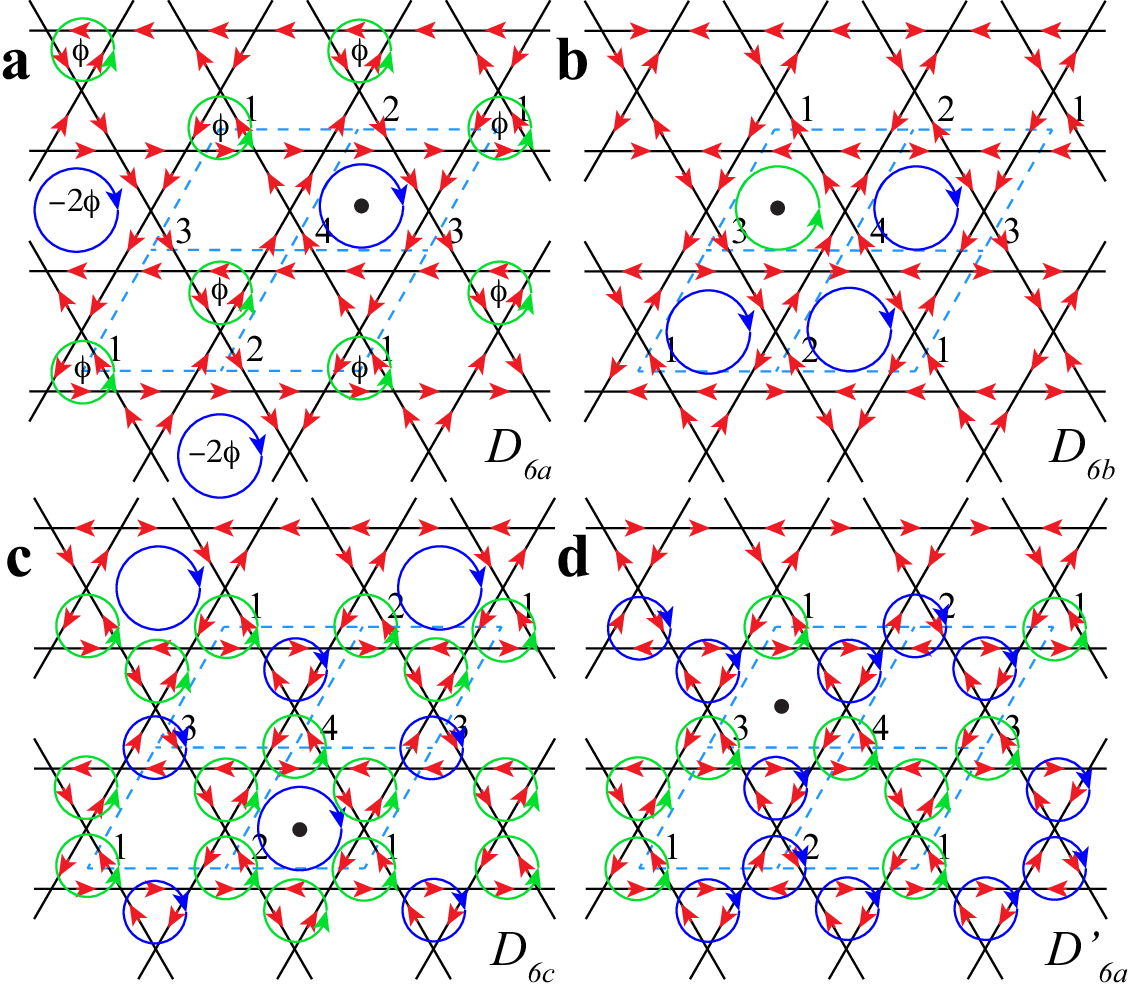}\caption{2*2 flux configurations  \textbf{a} $D_{6a}$ class with $-2\phi$ flux forming a triangle lattice and $\phi$ flux forming a honeycomb lattice. \textbf{b} $D_{6b}$ class  with one positive flux hexagon and three negative flux hexagons. \textbf{c} $D_{6c}$ class  with flipping the two diagonal triangle fluxes of three hexagons.  \textbf{d} $D'_{6a}$ class with three opposite flux triangle loops at each hexagon.  The inversion center of each class is also highlighted by black dots.
			\label{D6hs}}
	\end{center}
	\vskip-0.5cm
\end{figure}

The symmetry breaking orders with 2*2 unit cell are the most important configurations we focus on.  We search all possible configurations by the brute-force approach and project out the configurations violating the charge conservation rule. We find 122 flux phases with 2*2 unit cell. Among the 122 configurations, there are 18 classes, as summarized in Table.\ref{tab1} and shown in Fig. \ref{D6hs}-\ref{2*2,3}. $D_{6a}$ state is found to be the lowest energy state among all these flux states.

Similar to the previous discussion, there are 4 classes that only break the time reversal symmetry. Since the invariant subgroup of $D_{6h}$ can be either $C_{6h}$ or $D_{3h}$, these four classes can belong to different magnetic group $D_{6h}^{*}(C_{6h})$ or $D_{6h}^{*2}(D_{3h})$, as listed in Table.\ref{tab1}. For $D_{6h}^{*}(C_{6h})$ group, we find three classes $D_{6a}$, $D_{6b}$ and $D_{6c}$, as shown in Fig.\ref{D6hs}a-c. The $D_{6a}$ is the CFP state we proposed in the previous work \cite{feng}, where two $\phi$ fluxes form a honeycomb lattice with another $-2\phi$ form a triangle lattice. $D_{6a}$ class only contains two configurations related to time-reversal.

$D_{6b}$  class can be viewed as flipping one hexagon flux of the four four hexagon plaquettes by enlarging Nagaosa configuration to 2*2 unit cell, as shown in Fig.\ref{D6hs}b. 
The remaining $D_{6c}$ class is found by flipping the two diagonal triangle fluxes of three hexagons in  2*2 unit cell, as shown in Fig.\ref{D6hs}c. The low-energy effective model for $D_{6b}$  and $D_{6c}$ also correspond to $\Gamma_{2}-\Gamma_{5}+\Gamma_{7}$ (see SM Sec.1 for detail.\cite{app}). 
For $D_{6h}^{*2}(D_{3h})$ group, we flip three triangle fluxes of each hexagon plaquette, as shown in Fig.\ref{D6hs}d. Both $D_{6b,c}$ and $D'_{6a}$ only contains two configurations related by time-reversal.

\begin{table}[ht]
\begin{ruledtabular}
\caption{18 classes of flux phase in 2*2 unit cell and their symmetry groups. The number of configurations at each class is also listed at their brackets.}\label{tab1}
\begin{tabular}{c|ccccc}
Symmetry & Class Name \\\hline
$D_{6h}^{*}(C_{6h})$  & $D_{6a} (2) $,$D_{6b}(2)$,$D_{6c}(2)$                                   \\ \hline
$D_{6h}^{*2}(D_{3h})$ & $D_{6a}^{'}(2)$                                                    \\\hline
$D_{3h}^{*}(C_{3h})$  & $D_{3a}(4)$,$D_{3b}(4)$,$D_{3c}(4)$                                \\\hline
$D_{2h}^{*}(C_{2h})$  & $D_{2a}(6)$                                            \\\hline
$C_{2v}^{*}$          & $C_{2a}$(6),$C_{2b}$(12),$C_{2c}$(12),$C_{2d}$(12),\\
& $C_{2e}$(12),$C_{2f}$(12),$C_{2g}$(6),$C_{2h}$(6) \\\hline
$C_{2v}$              &  $C_{2}^{'}$  (6)                                            \\\hline
$C_{2}$               &  $C_{2}^{''}$  (12)       \\                     
\end{tabular}
\end{ruledtabular}
\end{table}

Besides above classes, other classes break both time reversal symmetry and point group symmetry. We find $D_{3h}^{*}(C_{3h})$ , $D_{2h}^{*}$($C_{2h}$), $C_{2v}^{*}$, $C_{2v}$ and $C_{2}$ groups. There are three classes belonging to $D_{3h}^{*}(C_{3h})$ as shown in Fig.\ref{2*2,2}(a)-(c). The category with 8 classes is $C_{2v}^{*}$ as shown in Fig.\ref{2*2,3}.(a)-(h). As for $D_{2h}^{*}$($C_{2h}$), $C_{2v}$, $C_{2}$, there are only one class in each category, as shown in Fig.\ref{2*2,2}(d), Fig.\ref{2*2,2}(e) and Fig.\ref{2*2,2}(f), respectively. The numbers of configurations for each class are summarized in Table.\ref{tab1}.

Note that the above flux constructing method is based on real space without considering the low-energy 3Q scattering at $5/4$ filling. Therefore, most of them are not relevant to AV$_3$Sb$_5$.  For example, the $C_{2g}$, $C_{2h}$, $C_{2'}$ and $C_{2''}$ configurations are gapless at vH filling.

\begin{figure*}
	\begin{center}
		\fig{7.0in}{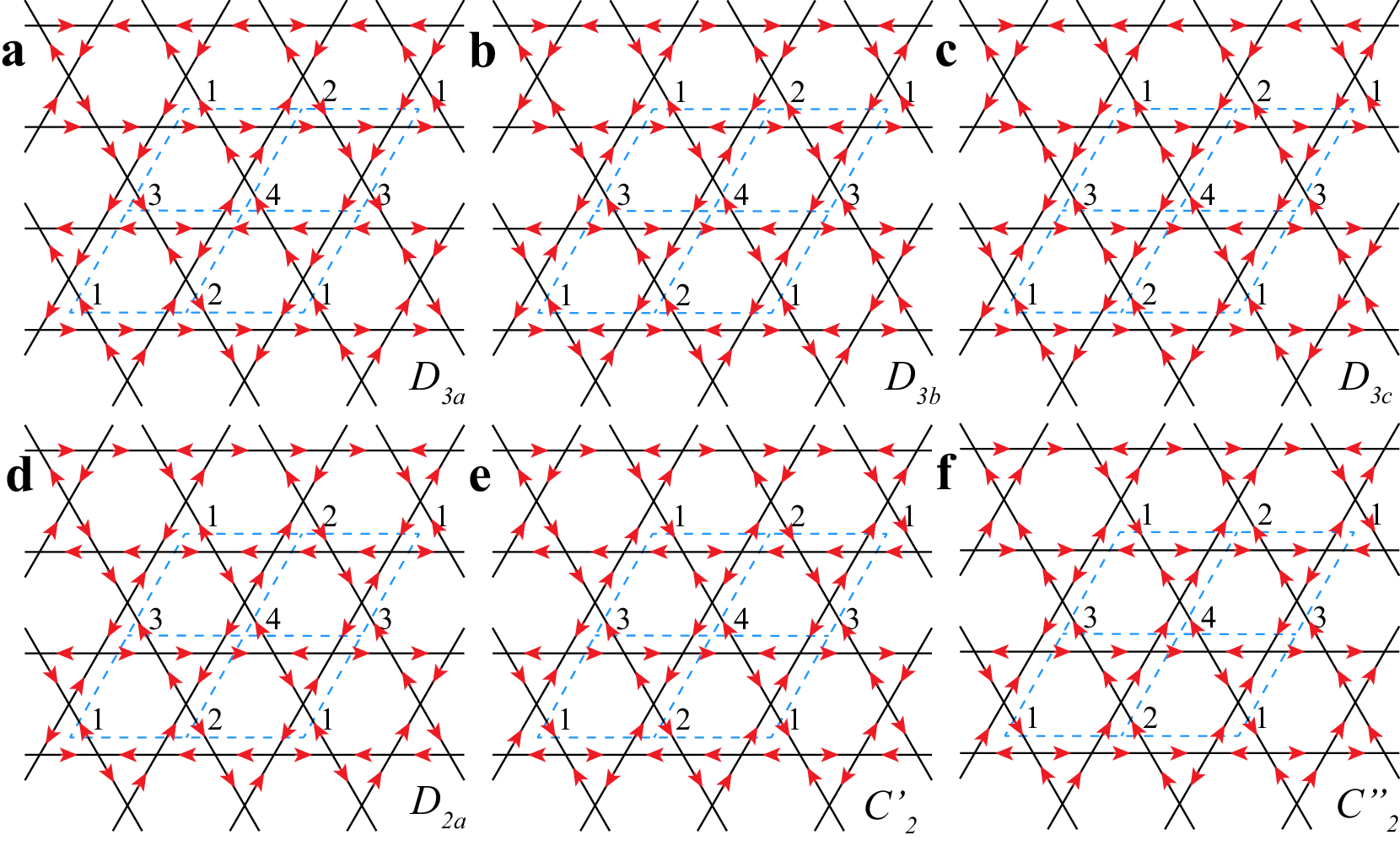}\caption{2*2 flux configuration  \textbf{a} $D_{3a}$ class.  \textbf{b} $D_{3b}$ class.  \textbf{c} $D_{3c}$ class.   \textbf{d} $D_{2a}$ class. \textbf{e} $C_{2}^{'}$  class. \textbf{f} $C_{2}''$ class .
			\label{2*2,2}}
	\end{center}
	\vskip-0.5cm
\end{figure*}

\begin{figure*}
	\begin{center}
		\fig{7.0in}{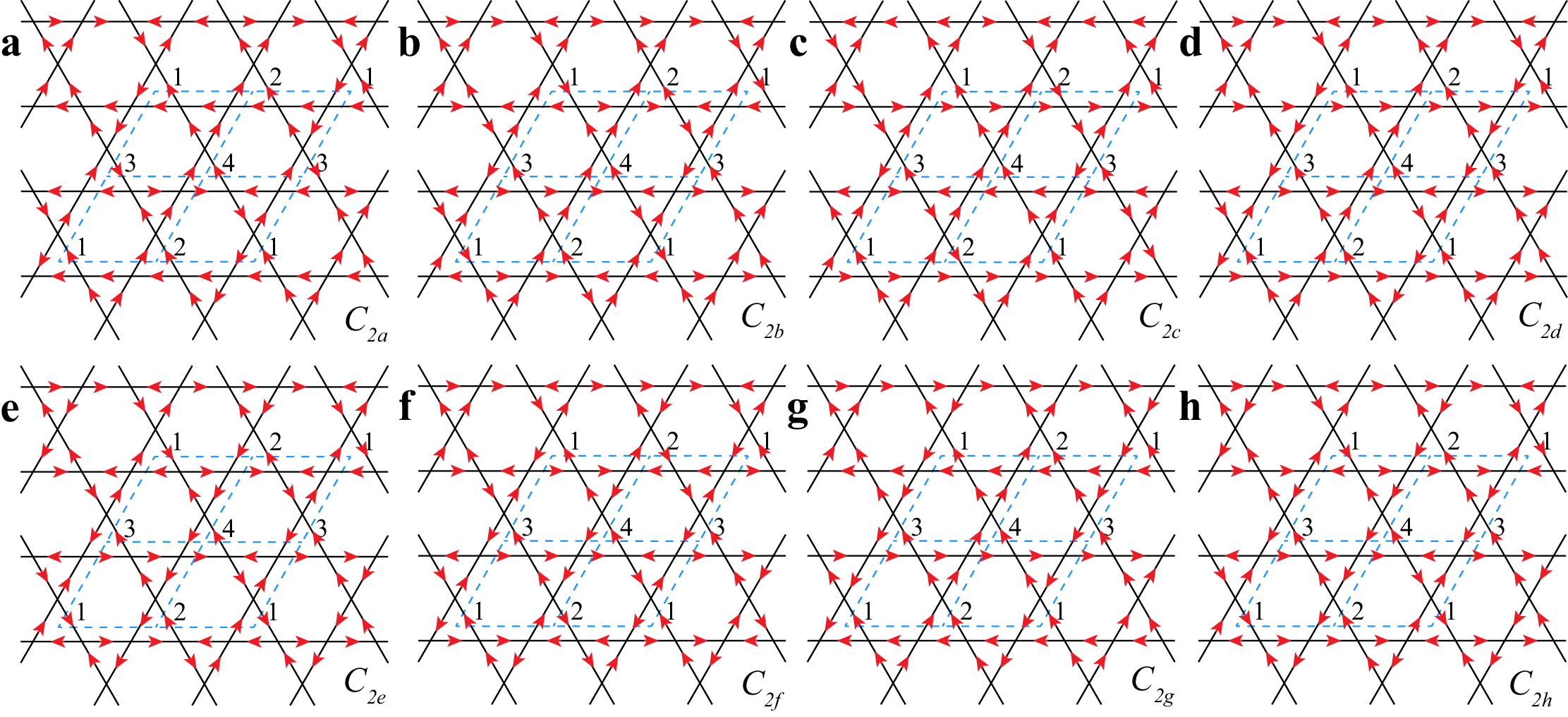}\caption{2*2 flux configurations  \textbf{a} $C_{2a}$ class.  \textbf{b} $C_{2b}$ class.  \textbf{c} $C_{2c}$ class.   \textbf{d} $C_{2d}$ class.   \textbf{e} $C_{2e}$ class.   \textbf{f} $C_{2f}$ class.  \textbf{g} $C_{2g}$ class.   \textbf{h} $C_{2h}$ class.     }
			\label{2*2,3}
	\end{center}
	\vskip-0.5cm
\end{figure*}

\subsection{1 * 2 configuration}
To complete our discussion on flux phases, we also list all the 1 * 2 configurations.
There are 17 flux phases in 1 by 2 unit cell, which only contain 8 classes (Table.\ref{tab2}), as shown in Fig. \ref{1*2,1}.  Since there are three translation directions in Kagome lattice, there are also another 2*17 flux phases by breaking translation symmetry in other two directions.

These 8 classes can be divided into 3 categories by symmetry, all of them break both time reversal symmetry and point-group symmetry. Similar to magnetic group $D_{6h}^{*}$, the invariant subgroup of the magnetic group $D_{2h}^{*}$ can be either $C_{2h}$ or $C_{2v}$. Thus, there are two kinds of magnetic groups. The classes shown in Fig.\ref{1*2,1} (a) and (b) belong to $D_{2h}^{*1}(C_{2h})$, while the class shown in Fig.\ref{1*2,1} (c) belongs to $D_{2h}^{*2}(C_{2v})$. Other 5 classes belong to $C_{2v}^{*}(C_{2})$ as shown in Fig.\ref{1*2,1} (d)-(h).
\begin{table}[ht]
\begin{ruledtabular}
\caption{8 classes of flux phase in 1*2 unit cell. The number of configurations at each class is also listed at their brackets.}
\begin{tabular}{c|cccc}\label{t12}
	Symmetry & Class Name \\\hline
 $D_{2h}^{*1}(C_{2h})$ & $D_{2a}(1)$,$D_{2b}(1)$                         \\ \hline
$D_{2h}^{*2}(C_{2v})$ & $D_{2a}^{'}(1)$                                \\ \hline
$C_{2v}^{*}(C_2)$         & $C_{2a}(2)$,$C_{2b}(2)$,$C_{2c}(2)$,$C_{2d}(4)$,$C_{2e}(4)$ \\ 
\end{tabular}
\end{ruledtabular}
\end{table}

\begin{figure*}
	\begin{center}
		\fig{7.0in}{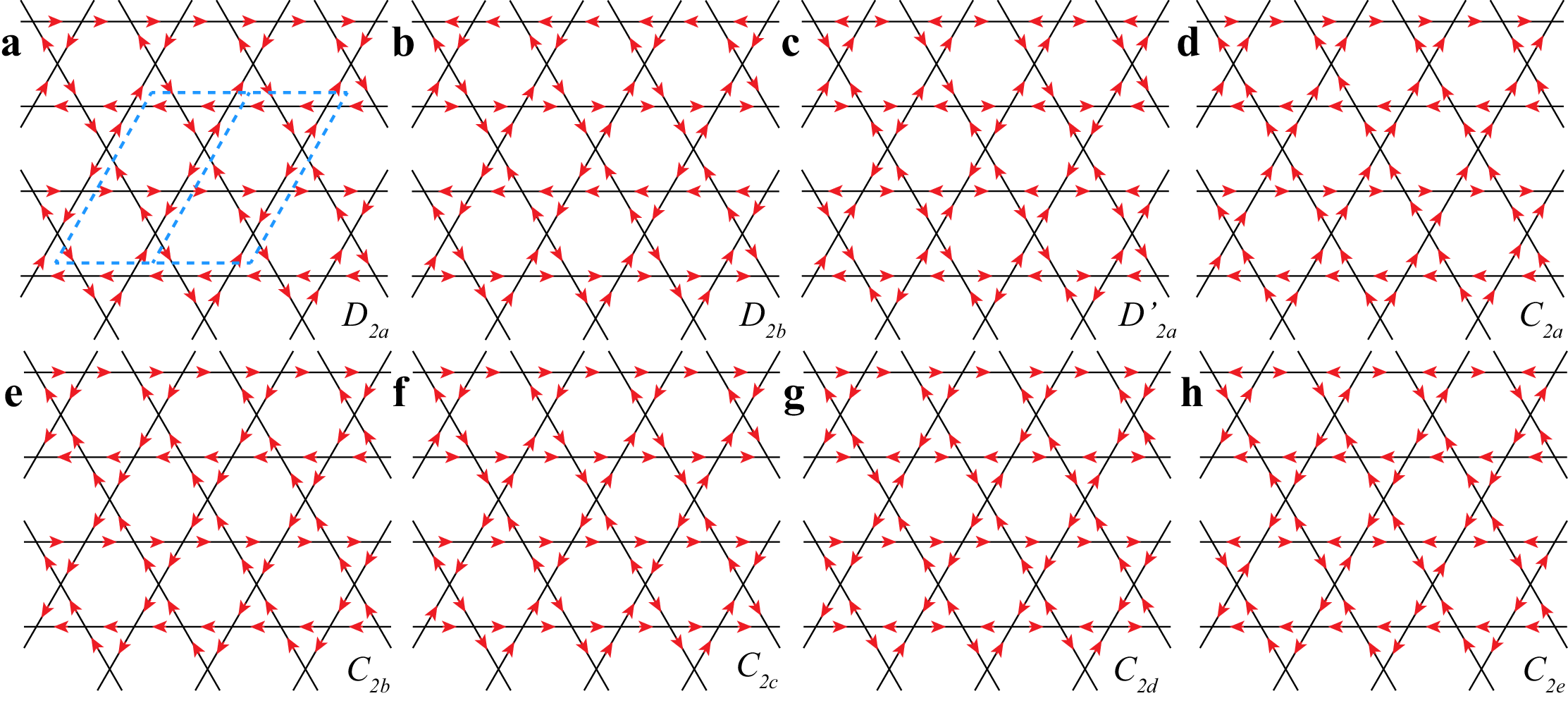}\caption{1*2 flux configurations  \textbf{a} $D_{2a}$  class.  \textbf{b} $D_{2b}$ class.  \textbf{c} $D'_{2a}$ class.  \textbf{d} $C_{2a}$ class. \textbf{e} $C_{2b}$  class. \textbf{f} $C_{2c}$ class. \textbf{g} $C_{2d}$ class. \textbf{h} $C_{2e}$ class.
			\label{1*2,1}}
	\end{center}
	\vskip-0.5cm
\end{figure*}

\subsection{$C_6$ symmetric CDW and CBO}
\begin{figure*}[ht]
	\begin{center}
		\fig{7.0in}{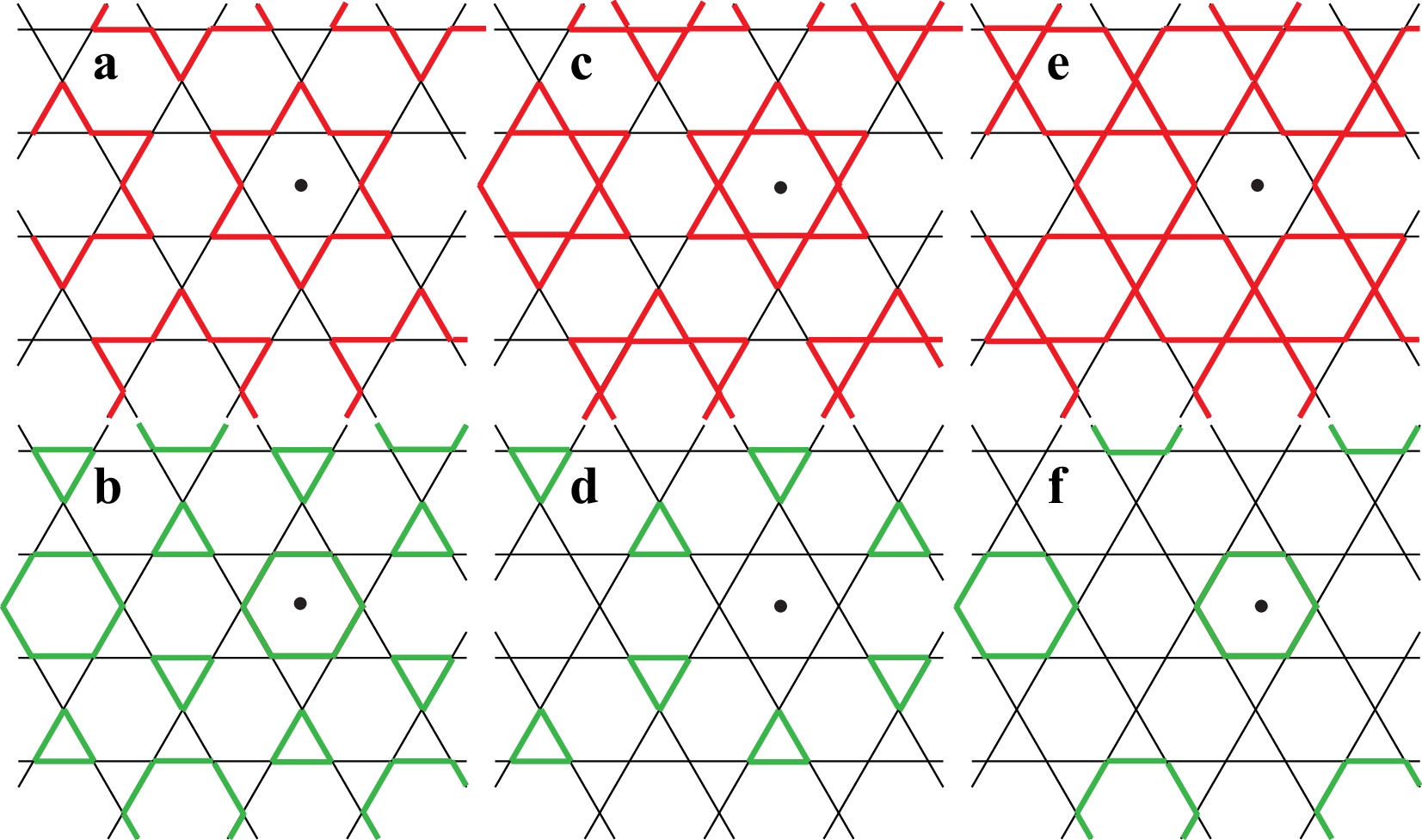}\caption{CBO configurations with $D_{6h}$ symmetry.  \textbf{a} $D_{6a}$ or ATH.  \textbf{b} $ID_{6a}$ or TrH.  \textbf{c} $D_{6b}$ or SoD.  \textbf{d} $ID_{6b}$ or ISoD. \textbf{e} $D_{6c}$. \textbf{f} $ID_{6c}$. 
			\label{D6h_CBO}}
	\end{center}
	\vskip-0.5cm
\end{figure*}

\begin{figure}[ht]
	\begin{center}
		\fig{3.4in}{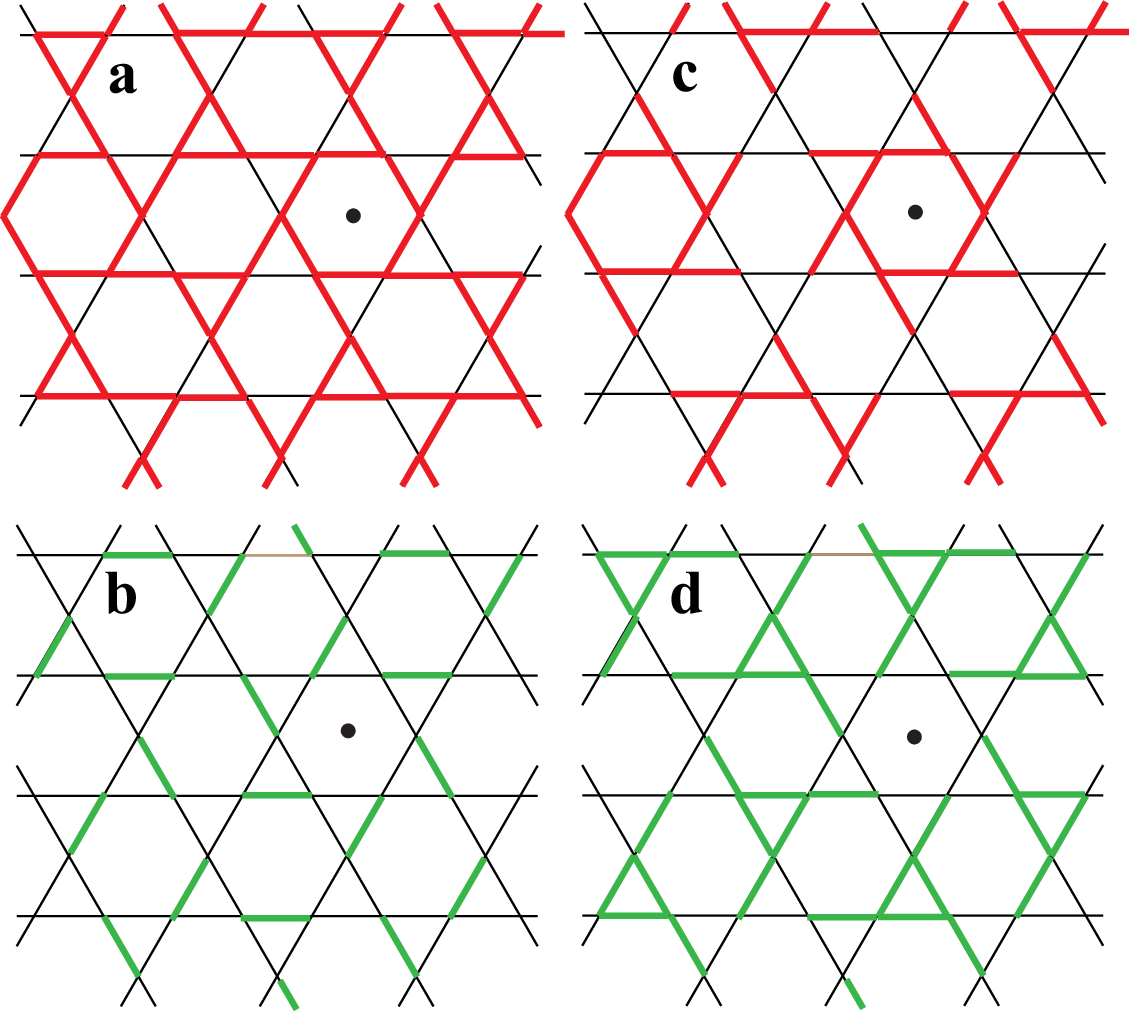}\caption{CBO configurations  with $C_{6h}$ symmetry.  \textbf{a} $C_{6a}$. \textbf{b} $IC_{6a}$.  \textbf{c} $C_{6b}$.  \textbf{d} $IC_{6b}$. 
			\label{C6h_CBO}}
	\end{center}
	\vskip-0.5cm
\end{figure}

Besides the above flux states, finding other charge density wave and charge bond order states is also an interesting task. Comparing to flux states, CDW could have $2^{12}$  possible configurations and CBO could have $2^{24}$ possible configurations within 2*2 unit cell, which seems to be an impossible work. However, we can use symmetry constraint to focus on high symmetry configurations. Utilizing the $C_6$ rotation symmetry constraint, we search all possible $C_6$ symmetric CDWs and CBOs. For CDW state, we only find 2 states with positive or negative charge vCDW-b configuration. 
For CBO states, we find 6 CBOs with $D_{6h}$ point group symmetry and 4 CBOs with $C_{6h}$ point group symmetry, as shown in Fig.\ref{D6h_CBO} and Fig.\ref{C6h_CBO}. In Fig.\ref{D6h_CBO}, the up panel  states are labeled as $D_{6a}$, $D_{6b}$ and $D_{6c}$ while the down panel states are inverse configurations of the up panel labeled as $ID_{6a}$, $ID_{6b}$ and $ID_{6c}$. The $D_{6a}$ state is ATH, the $ID_{6a}$ is TrH and the $D_{6b}$ is SoD as defined above. Notice that $ID_{6b}$ is the "Inverse Star of David" (ISoD) rather than $ID_{6a}$. Similarly, we also label the $C_{6h}$ symmetry configurations as $C_{6a}$, $C_{6b}$ and $IC_{6a}$, $IC_{6b}$, as shown in Fig.\ref{C6h_CBO}.

\section{Discussion and Summary}
In summary, the time-reversal symmetry breaking CDW found in AV$_3$Sb$_5$ provides a new platform to investigate correlation and topology.  By analyzing the dominated van-Hove points around the Fermi surface and corresponding symmetry properties,  the low-energy effective theory for Kagome lattice at vH filling can be constructed. All symmetry breaking states of this vH low-energy effective theory can be found and classified according to the point group. The relations between low-energy symmetry breaking states to the physical orders in real space can be fully established.  The above study based on a single orbital model on Kagome lattice can be straightforwardly generalized to multi-orbital cases. The dominated time-reversal breaking channel is always the flux phase. The full flux configurations that satisfy the charge conservation rule include 183 flux phases in kagome lattice within 2*2 unit cell. Especially, we list all 3, 18, 8 classes which are independent in symmetry belonging to 1*1, 2*2, and 1*2 unit cell in Table.\ref{tab2}, \ref{tab1} and \ref{t12}, respectively. The symmetry classifications and relations to low-energy effective theory of these flux phases are obtained. 
 All these findings give rise to complete analysis and new understandings of flux phases in Kagome lattice.

Note that, when finalizing this work, several theoretical works starting from low-energy effective theory appeared \cite{balents,nandkishore}. Ref. \cite{nandkishore} analyzed the real and imaginary CDW at vH singularity on the hexagonal lattices from a phenomenological Ginzburg-Landau theory. Ref. \cite{balents} studied the electronic instabilities of Kagome lattice using parquet renormalization group and corresponding Landau theory.

We thank Xi Dai, Li Yu, Hu Miao and Ziqiang Wang for useful discussions. This work is supported by the Ministry of Science and Technology  (Grant
No. 2017YFA0303100), National Science Foundation of China (Grant No. NSFC-11888101), and the Strategic Priority Research Program of Chinese Academy of Sciences (Grant
No. XDB28000000). K.J. acknowledges support from the start-up grant of IOP-CAS.


\begin{thebibliography}{99}
	
	\bibitem{haldane}
	F. D. M. Haldane, Phys. Rev. Lett. {\bf{61}}, 2015 (1988). 
	
	\bibitem{affleck}
	Ian Affleck and J. Brad Marston, Phys. Rev. B {\bf37}, 3774(R) (1998).
	
	\bibitem{lee}
	Menke U. Ubbens and Patrick A. Lee, Phys. Rev. B {\bf46}, 8434 (1992).
	
	\bibitem{doping}
	Patrick A. Lee, Naoto Nagaosa, and Xiao-Gang Wen, Rev. Mod. Phys. {\bf78}, 17 (2006).
	
	\bibitem{varma}
	C. M. Varma, Phys. Rev. B {\bf55}, 14554(1997).
	
	\bibitem{chakravarty}
	Sudip Chakravarty, R. B. Laughlin, Dirk K. Morr, and Chetan Nayak, Phys. Rev. B {\bf63}, 094503 (2001).
	
	\bibitem{varma97}
	C. M. Varma, Phys. Rev. B {\bf55}, 14554(1997).
	\bibitem{varma06}
	C. M. Varma, Phys.Rev.B {\bf73}, 155113 (2006).
	
	\bibitem{norman}
	M. R. Norman ,D. Pines, and C. Kallin,  Adv. Phys. {\bf54}, 715 (2005).
	
	\bibitem{keimer}
	B. Keimer, S. A. Kivelson, M. R. Norman, S. Uchida, and J. Zaanen, Nature {\bf518}, 179 (2015).
	
	\bibitem{Venderbos}
  J. W. F.Venderbos, Phys. Rev. B {\bf93},115107 (2016)
	
	   \bibitem{yxzhang}
   Y.X.Zhang, H.M.Guo, R.T.Scalettar, Phys.Rev.B.{\bf101},205139 (2020)
   \bibitem{nayak}
   Chetan Nayak, Phys. Rev. B {\bf62}, 4880(2000)
   \bibitem{chang}
   Chia-Chen Chang and Richard T. Scalettar, Phys. Rev. Lett. {\bf109}, 026404 (2012)
   \bibitem{hayami}
   Satoru Hayami, Yuki Yanagi, Hiroaki Kusunose, and Yukitoshi Motome, Phys. Rev. Lett. {\bf122}, 147602 (2019).
   \bibitem{liu}
   Jianpeng Liu, Se Young Park, Kevin F. Garrity, and David Vanderbilt, Phys. Rev. Lett. {\bf117}, 257201 (2016).
   \bibitem{luo}
   Zhu-Xi Luo, Cenke Xu, and Chao-Ming Jian, Phys. Rev. B {\bf104}, 035136 (2021).
   \bibitem{Shaik}
   N. E. Shaik, B. Dalla Piazza, D. A. Ivanov, and H. M. Rønnow, Phys. Rev. B {\bf102}, 214413(2020).
    
	\bibitem{ortiz19}
	Brenden R. Ortiz, Lídia C. Gomes, Jennifer R. Morey, Michal Winiarski, Mitchell Bordelon, John S. Mangum, Iain W. H. Oswald, Jose A. Rodriguez-Rivera, James R. Neilson, Stephen D. Wilson, Elif Ertekin, Tyrel M. McQueen, and Eric S. Toberer, Phys. Rev. Materials. {\bf3}, 094407 (2019).
	
	\bibitem{ortiz20}
	Brenden R. Ortiz, Samuel M.L. Teicher, Yong Hu, Julia L. Zuo, Paul M. Sarte, Emily C. Schueller, A.M. Milinda Abeykoon, Matthew J. Krogstad, Stephan Rosenkranz, Raymond Osborn, Ram Seshadri, Leon Balents, Junfeng He, and Stephen D. Wilson, Phys. Rev. Lett. {\bf125}, 247002 (2020). 
	
	\bibitem{topocdw}
	Yu-Xiao Jiang, Jia-Xin Yin, M. Michael Denner, Nana Shumiya, Brenden R. Ortiz, Gang Xu, Zurab Guguchia, Junyi He, Md Shafayat Hossain, Xiaoxiong Liu, Jacob Ruff, Linus Kautzsch, Songtian S. Zhang, Guoqing Chang, Ilya Belopolski, Qi Zhang, Tyler A. Cochran, Daniel Multer, Maksim Litskevich, Zi-Jia Cheng, Xian P. Yang, Ziqiang Wang, Ronny Thomale, Titus Neupert, Stephen D. Wilson, M. Zahid Hasan, Nat. Mater. {\bf20}, 1353–1357 (2021). 
	
	\bibitem{hall20}
	Shuo-Ying Yang, Yaojia Wang, Brenden R Ortiz, Defa Liu, Jacob Gayles, Elena Derunova, Rafael Gonzalez-Hernandez, Libor Šmejkal, Yulin Chen, Stuart S P Parkin, Stephen D Wilson, Eric S Toberer, Tyrel McQueen, Mazhar N Ali, Sci. Adv. {\bf6}, eabb6003 (2020).
	
	\bibitem{xhchen}
	F. H. Yu, T. Wu, Z. Y. Wang, B. Lei, W. Z. Zhuo, J. J. Ying, X. H. Chen, Phys. Rev. B {\bf104}, L041103.
	
	\bibitem{musr}
	Li Yu, Chennan Wang, Yuhang Zhang, Mathias Sander, Shunli Ni, Zouyouwei Lu, Sheng Ma, Zhengguo Wang, Zhen Zhao, Hui Chen, Kun Jiang, Yan Zhang, Haitao Yang, Fang Zhou, Xiaoli Dong, Steven L. Johnson, Michael J. Graf, Jiangping Hu, Hong-Jun Gao, Zhongxian Zhao. arXiv:2107.10714.
	
	\bibitem{feng}
	Xilin Feng, Kun Jiang, Ziqiang Wang, Jiangping Hu, Science Bulletin {\bf66(14)}, 1384-1388 (2021).
	
    
	\bibitem{ronny}
	M. Michael Denner and Ronny Thomale and Titus Neupert, arxiv:2103.14045. 
		
	\bibitem{nandkishore}
	Yu-Ping Lin and Rahul M. Nandkishore, Phys. Rev. B {\bf104}, 045122.
	
	\bibitem{balents}
	Takamori Park, Mengxing Ye, Leon Balents, Phys. Rev. B {\bf104}, 035142 (2021).
	
 \bibitem{gu}
    Yuhao Gu, Yi Zhang, Xilin Feng, Kun Jiang, Jiangping Hu, 	arXiv:2108.04703.
    
    \bibitem{kun}
    Kun Jiang, Tao Wu, Jia-Xin Yin, Zhenyu Wang, M. Zahid Hasan, Stephen D. Wilson, Xianhui Chen, Jiangping Hu, arXiv:2109.10809.
    
    	\bibitem{app}
	Supplement Information of this paper. Sec.1 Basis transformation and relations between 3Q scattering and 3Q pattern in real space. Sec.2 Gauge transformation and translation operations.
 \bibitem{martin}
Ivar Martin and C. D. Batista, Phys. Rev. Lett. {\bf101}, 156402 (2008).

\bibitem{taoli}
Tao Li, EPL {\bf97} 37001 (2012).

\bibitem{motome}
Satoru Hayami and Yukitoshi Motome, Phys. Rev. B {\bf90}, 060402(R) (2014).

\bibitem{jxli}
Shun-Li Yu and Jian-Xin Li, Phys. Rev. B {\bf85}, 144402 (2012).

\bibitem{chubukov}
R. Nandkishore, L. Levitov, and A. Chubukov, Nat. Phys. {\bf8}, 158 (2012).

\bibitem{qhwang12}
Wan-Sheng Wang, Yuan-Yuan Xiang, Qiang-Hua Wang, Fa Wang, Fan Yang, and Dung-Hai Lee, Phys. Rev. B {\bf85}, 035414 (2012).

\bibitem{thomale12}
Maximilian L. Kiesel, Christian Platt, Werner Hanke, Dmitry A. Abanin, and Ronny Thomale, Phys. Rev. B {\bf86}, 020507(R) (2012).

\bibitem{qhwang13}
Wan-Sheng Wang, Zheng-Zhao Li, Yuan-Yuan Xiang, and Qiang-Hua Wang, Phys. Rev. B {\bf87}, 115135 (2013).

\bibitem{thomale13}
Maximilian L. Kiesel, Christian Platt, and Ronny Thomale, Phys. Rev. Lett. {\bf110}, 126405 (2013).
	
	\bibitem{wilson}
	Brenden R. Ortiz, Samuel M. L. Teicher, Linus Kautzsch, Paul M. Sarte, Jacob P. C. Ruff, Ram Seshadri, Stephen D. Wilson,  arXiv: 2104.07230.
	
	\bibitem{tsirlin}
	E. Uykur, B. Ortiz, S. Wilson, M. Dressel, and A. Tsirlin, arXiv:2103.07912.

\bibitem{yan}	
	H. Tan, Y. Liu, Z. Wang, and B. Yan, Phys. Rev. Lett. {\bf127}, 046401 (2021)
	\bibitem{nagaosa}
	Kenya Ohgushi, Shuichi Murakami, and Naoto Nagaosa, Phys. Rev. B {\bf62}, R6065(R) (2000).
	

\end{thebibliography}
\end{document}